# OMNIMAMBA4D: SPATIO-TEMPORAL MAMBA FOR LONGITUDINAL CT LESION SEGMENTATION

*Justin Namuk Kim[1*], Yiqiao Liu[2], Rajath Soans[2], Keith Persson[2], Sarah Halek[2], Michal Tomaszewski[2], Jianda Yuan[2], Gregory Goldmacher[2], Antong Chen[2]*

[1] Department of Biomedical Engineering, Case Western Reserve University, Cleveland, OH, USA
[2] Merck & Co., Inc., Rahway, NJ, USA

## ABSTRACT

Accurate segmentation of longitudinal CT scans is important for monitoring tumor progression and evaluating treatment responses. However, existing 3D segmentation models solely focus on spatial information. To address this gap, we propose OmniMamba4D, a novel segmentation model designed for 4D medical images (3D images over time). OmniMamba4D utilizes a *spatio-temporal tetra-orientated Mamba* block to effectively capture both spatial and temporal features. Unlike traditional 3D models, which analyze single-time points, OmniMamba4D processes 4D CT data, providing comprehensive spatio-temporal information on lesion progression. Evaluated on an internal dataset comprising of 3,252 CT scans, OmniMamba4D achieves a competitive Dice score of 0.682, comparable to state-of-the-arts (SOTA) models, while maintaining computational efficiency and better detecting disappeared lesions. This work demonstrates a new framework to leverage spatio-temporal information for longitudinal CT lesion segmentation.

## 1. INTRODUCTION

Traditional endpoints in oncology trials involve measuring longitudinal tumor burden changes using Response Evaluation Criteria in Solid Tumors (RECIST). While RECIST-based endpoints are strong predictors of clinical outcome in large cohorts, they do not correlate well with survival in small cohorts [1]. To improve prognostic prediction, active research focuses on radiomics-based approaches [2] and model-informed methods [3]. Central to these approaches is accurate 3D lesion segmentation in longitudinal computed tomography (CT) scans across all lesions, which is essential for measuring total tumor burden changes, monitoring tumor progression, and evaluating treatment responses. However, manual annotation of longitudinal CT scans is time-consuming, necessitating automated deep learning solutions.

Convolutional Neural Network (CNN)-based models, such as nnU-Net [4], are widely used for medical imaging tasks due to their ability to capture local spatial features. However, CNNs struggles to model long range dependencies due to their limited receptive field. Transformers [5,6] have shown promise in addressing this limitation, by utilizing self-attention mechanism to capture global features, gaining significant traction in 3D medical imaging tasks [7,8]. Despite their advantages, transformers have quadratic computational complexity, making them less suitable for 4D medical datasets with large input token lengths.

To address the challenges of 4D medical segmentation, a couple methods have been proposed to date. Yao et al [9] introduced a framework, integrating fully connected networks for spatial modeling with long short-term memory networks for temporal modeling. Sun et al [10] proposed utilizing true 4D convolutions, enabling direct joint learning of spatial and temporal features directly using convolutions. While these approaches provide initial solutions, further innovation is necessary to efficiently and effectively handle large 4D medical datasets.

The Mamba model [11], derived from State Space Model (SSM), offers an efficient alternative for capturing long-range dependencies. Structure state space sequence models (S4) [12] excel at handling such dependencies, with Mamba improving upon S4 by incorporating a selective mechanism that adaptively focuses on relevant information in an input-dependent manner. Mamba-based models –with its selective mechanism and hardware-aware algorithms [11]– outperform transformers in both performance and efficiency in computer vision [13] tasks, highlighting their potential of capturing long-range dependencies in medical images.

Several studies have explored Mamba-based models in 2D and 3D medical images. U-MAMBA [14] embeds Mamba blocks into the encoder or bottleneck, while Vivim [15] and SegMAMBA [16] apply spatio-temporal learning for 2D images across multiple frames, and tri-oriented Mamba for 3D images, respectively. Vision MAMBA [17] uses bidirectional Mamba blocks, and nnMamba [18] integrates SSM with CNNs, leveraging channel and spatial feature interaction. LKM-UNet [19] employs large-kernel, pixel-level, and patch-level bidirectional Mamba. Despite their successes, no prior work has investigated Mamba-based spatio-temporal segmentation networks for 4D medical images, such as longitudinal CT scans, across multiple time points, or compared these approaches with 3D-based models.

In this paper, we propose OmniMamba4D, a novel approach that leverages spatio-temporal Mamba blocks for segmentation of various types of lesions in longitudinal CT scans acquired at multiple time points. Based on a pipeline to propagate the baseline 3D lesion annotations from radiologists to later scans, OmniMamba4D expands on the tri-oriented Mamba-based encoder from SegMAMBA [16] into a spatio-temporal tetra-oriented Mamba encoder. The tetra-oriented Mamba encoder introduces a novel approach to simultaneously capture spatial and temporal dependencies. By efficiently extracting multi-scale spatio-temporal features, OmniMamba4D aims to achieve accurate and robust lesion segmentation across longitudinal CT data. Extensive experiments are conducted on a large internal dataset comprising 3,252 CT scans with expert annotations of lesions in various organs (e.g., lung, lymph node, and liver). We perform evaluations to compare the performance of 3D and 4D segmentation models.

## 2. METHODS

In this section, we present the overall architecture of OmniMamba4D, along with its core modules, which are designed to process high-dimensional spatio-temporal features from longitudinal CT images.

### 2.1 OmniMamba4D

To accommodate 4D medical image datasets, we extended the conventional 5D tensor input dimensions *(B,C,D,H,W)* in PyTorch to 6D tensors with the shape *(B,C,T,D,H,W)*, where *B* is the batch size, *C* is the channel size, *T* is the number of time-points, *H* is the height, *W* is the width, and *D* is the depth. During convolution, these tensors are reshaped from *(B,C,T,D,H,W)* to *(B×T,C,D,H,W)* to enable processing as 5D tensors, effectively treating each time point as a separate batch. While convolutional features do not utilize temporal features, our proposed TetraMamba Block is designed to learn temporal features efficiently.

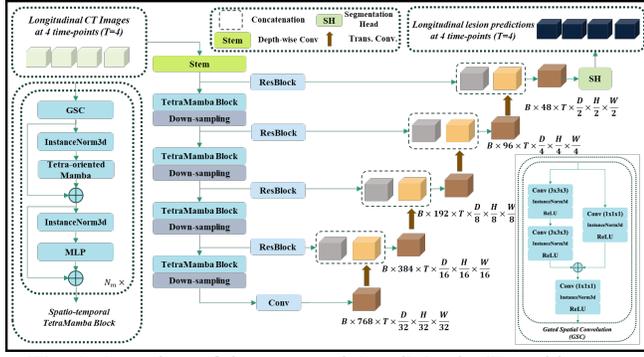

**Fig. 1.** Overview of the proposed OmniMamba4D architecture

*2.1.1 Encoder*
The OmniMamba4D architecture is designed to process the large sequence lengths inherent in 4D medical images. For example, an input of (4,64,64,64) with the dimension order of (*T,D,H,W*) results in a sequence length of 1,048,576, imposing a significant computational burden for transformer architectures. Our method addresses this challenge through its design, which incorporates multi-scale, spatio-temporal feature modeling with computational efficiency.

The encoder begins with a stem layer that includes a depth-wise convolution with a large kernel size of 7×7×7, padding of 3×3×3, and a stride of 2×2×2. Large-kernel convolution is used to capture broader contextual information and to increase the receptive field.

Given a 4D input volume $I \in \mathbb{R}^{B \times C \times T \times D \times H \times W}$, where C is the number of input channels, the stem layer produces the first resolution feature $\mathcal{F}_0 \in \mathbb{R}^{B \times 48 \times T \times \frac{D}{2} \times \frac{H}{2} \times \frac{W}{2}}$. This feature, $\mathcal{F}_0$, is flattened and reshaped into $(B, L_0, C_0)$, where $L_0$ is a length of sequence, computed as $T \times \frac{D_0}{2} \times \frac{H_0}{2} \times \frac{W_0}{2}$. The reshaped $\mathcal{F}_0$ is then processed through TetraMamba Block.

$$\hat{z}_n^l = GSC(z_n^l), \qquad \bar{z}_n^l = TetraMamba\left(IN(\hat{z}_n^l)\right) + \hat{z}_n^l,$$
$$z_n^{l+1} = MLP\left(IN(\bar{z}_n^l)\right) + \bar{z}_n^l \qquad (1)$$

For the $n^{th}$ TetraMamba block, the processing is represented in equation (1), where $GSC$ and $TetraMamba$ represent gated spatial convolution and spatio-temporal tetra-orientated Mamba, respectively. $IN$ and $MLP$ represent instance normalization and multilayer-perceptron to enhance the feature representation.

For the Tetra-oriented Mamba, we build upon the tri-orientated Mamba from SegMamba [16], which was designed to capture inter-slice feature interactions along with bidirectional feature interactions in 3D medical imaging, by extending it to additionally model temporal interactions in 4D datasets. While SegMamba efficiently captures spatial dependencies across slices in 3D images, Tetra-orientated Mamba enhances this by incorporating temporal feature dependencies simultaneously as shown in equation (2).

$$TetraMamba(z) = M(z_f) + M(z_r) + M(z_s) + M(z_t) \qquad (2)$$

To model global information in high-dimensional spatio-temporal data, Tetra-oriented Mamba computes the feature dependencies from four distinct directions: forward, reverse, inter-slice, and temporal. The inter-slice direction inherited from SegMamba, models interactions across spatial slices in depth dimensions, while our additional temporal direction models relationships across different time points. By flattening the 4D input features into four sequences, we perform features interaction, obtaining fused spatio-temporal features. In this setup, *M* represents the Mamba layer responsible for capturing global information within a sequence, with *f* denoting forward direction, *r* reverse direction, *s* inter-slice direction, and *t* temporal direction.

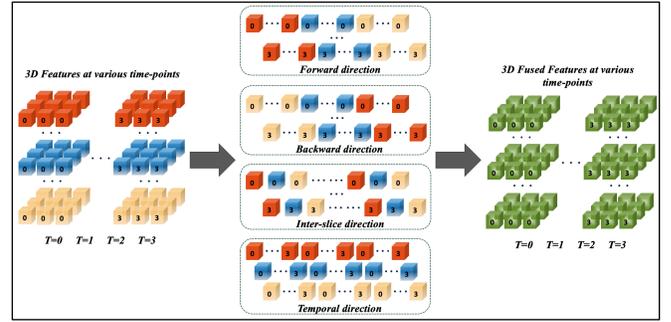

**Fig. 2.** Illustration of Tetra-orientated Mamba, processing spatial and temporal information in longitudinal CT images.

*2.1.2 Decoder*
The decoder in OmniMamba4D is CNN-based, inspired by established segmentation frameworks [16,20]. This architecture is designed to reconstruct the spatio-temporal features extracted during the encoding process, enabling high-resolution segmentation results. Skip connections facilitate feature re-use, enhancing the model's ability to capture fine-grained details. The segmentation head of decoder produces *T* segmentations, corresponding to each of the input time points.

### 2.2 Datasets

We evaluated our approach on an internal clinical dataset of 3,252 CT scans, which includes 813 unique lesions imaged at four time points with ground truth masks from radiologists across the four time points. The internal dataset comprised 400 lung lesions, 289 lymph node lesions, 52 adrenal lesions, 49 liver lesions, 14 soft tissue lesions, 6 abdominal lesions, and 3 bone lesions. The input to our pipeline is the baseline lesion annotation, and our proposed pipeline start with propagating ground truth lesion masks from the baseline (screening) scan (*T=0*) to follow-up scans (*T=1,2,3*), using an atlas-based segmentation via rigid and non-rigid registration. The registration quality has been reviewed by experts and confirmed suitable for the next step. Propagated lesion masks provide approximated location of lesions in follow-up scans, then both, screening and follow-up lesions were cropped with a margin twice their longest diameter and resized to 64×64×64 volumes. The dataset was randomly partitioned into 600 training, 100 validation, and 113 test lesions for the 4D models. The same dataset was used to train 3D segmentation models, using all four time-points. For the purpose of comparison, 3D models used input sizes of 64×64×64 for single time-point images, and OmniMamba4D models used input sizes of 4×64×64×64, representing four time-points. There is no data leakage across training, validation, and test set.

**Table 1. Dataset split for experiments.**

| Methods | Training | Validation | Testing |
|---|---|---|---|
| 3D | 2400 | 400 | 452 |
| 4D | 600 | 100 | 113 |

### 2.3 Pre-processing

Lesions were cropped based on their longest diameter $d$ measured in the xy plane, resulting in volumes of $2d \times 2d \times 2d$. The intensities in the range of -1024 HU to 2000 HU were normalized to 0-1 and the images were resized to size 64×64×64. Data augmentations including random scaling in the range of (-0.1, 0.1) with probability of 0.5, random rotations in the range of (-10°, 10°) with probability of 0.5, random flipping with probability of 0.3, and random intensity shifts in the range of (-0.1, 0.1) with probability of 0.5, were applied across 3D and 4D methods.

### 2.4 Model Comparison

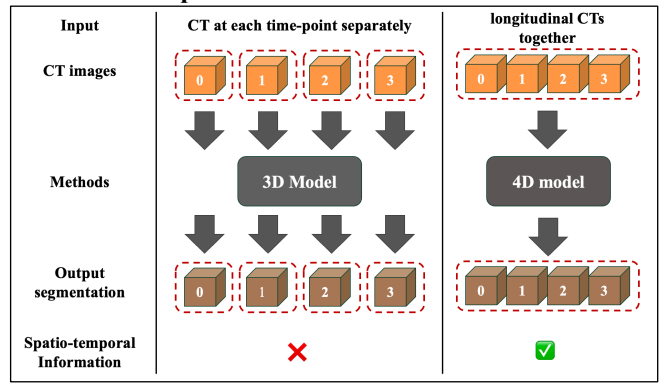

**Fig. 4.** Pictorial representation of difference in 3D and 4D methods

In 3D models, the networks were trained on 3D CT images from individual time points, predicting each time point independently. These models rely solely on spatial information from individual CT volumes. In contrast, 4D models were trained with 3D CT images across four time points from each lesion as 4D inputs, predicting segmentation for all four time-points at once. This allows the model to not only capture spatial information but also learn temporal dependencies across the different time points.

### 2.4 Implementation Details

We implemented our models using Pytorch 2.01 CUDA 11.8. Both 3D and 4D models are trained using two NVIDIA A100 GPUs. The models are optimized using AdamW optimizer with a learning rate of 0.001 and a weight decay of 0.05. The batch size is set to two and the number of epochs for training is empirically set to 200 for 3D models and 100 for 4D models according to the speed of convergence. For 3D models, maximum epoch was set to 200 as they required more epochs to converge at validation Dice. The Dice loss is used for training, and the model weights corresponding to the best validation Dice score are saved for evaluation on the test set.

## 3. PERFORMANCE EVALUATION

We evaluated the performance of OmniMamba4D against several state-of-the-art (SOTA) models using the Dice scores, which is a widely accepted metric for segmentation tasks. Table 2 summarizes the core modules used in each of the compared SOTA models, as well as the part of their integration into the network architecture.

**Table 2. SOTA models in medical image segmentation.**

| | Methods | Core Modules | Integration |
|---|---|---|---|
| | nnU-Netv2 | Convolution | Encoder |
| | LKMUNet | Large-kernel convolution with Bidirectional Mamba | Encoder |
| 3D | UMamba_bot | Uni-directional Mamba | Bottleneck |
| | UMamba_enc | | Encoder |
| | U-xLSTM_bot | xLSTM | Bottleneck |
| | U-xLSTM_enc | | Encoder |
| | SegFormer3D | MixViT | Encoder |
| | SwinUNETRv2 | SwinTransformer | Encoder |
| | SegMamba | Tri-orientated Mamba | Encoder |
| 4D | OmniMamba4D | Tetra Mamba | Encoder |

## 4. RESULTS

**Table 3.** Quantitative comparison on internal dataset. The bold and underlined Dice show the best and second-best performance.

| | Methods | Parameters (M) | Validation Dice ↑ | Test Dice ↑ |
|---|---|---|---|---|
| | nnU-Netv2 | 16.10 | 64.117 | 63.976 |
| | LKMUNet | 16.10 | 57.855 | 59.550 |
| | UMamba_bot | 21.77 | 67.270 | 67.738 |
| | UMamba_enc | 21.76 | 66.094 | 66.594 |
| 3D | U-xLSTM_bot | 21.73 | 64.771 | 65.005 |
| | U-xLSTM_enc | 22.56 | 64.880 | 65.001 |
| | SegFormer3D | 4.49 | 64.987 | 65.850 |
| | SwinUNETRv2 | 72.76 | <u>68.220</u> | <u>68.864</u> |
| | SegMamba | 67.36 | **68.992** | **69.890** |
| 4D | OmniMamba4D | 67.36 | 66.669 | 68.227 |

### 4.1 Comparison with the SOTA models

We compared OmniMamba4D with several SOTA segmentation models including SegMAMBA [16], SwinUNETRv2 [20], SegFormer3D [21], U-xLSTM [22]. The SOTA models are chosen based on segmentation performance on various medical image datasets as of date the study was conducted. The Dice score was used to evaluate performance on the validation and test set, as shown in Table 3. The number of model parameters are also shown.

Among the 3D models, SegMamba achieved the highest Dice score on the test set of 69.890, followed by SwinUNETRv2 with 68.864. Our OmniMamba4D model achieved a Dice score of 68.227, outperforming several 3D models including nnU-Netv2 [23], LKM-UNet, U-Mamba, U-xLSTM, and SegFormer3D. While OmniMamba4D processes *four times* the input sequence length of 3D models, it demonstrated competitive performance with relatively fewer parameters than SwinUNETRv2 and effectively capturing both spatial and temporal information.

## 4.2 Visual Comparison of SegMamba (3D) and OmniMamba4D (4D)

We compared the segmentation results of the top-performing models– SegMamba (3D) and OmniMamba4D (4D). We showed six example lesions in lymph node, lung, and liver in Figure 5 with OmniMamba4D in yellow contour, SegMamba in blue contour, and ground truth in red contour. The second lymph node lesion example (second row from Fig. 5) shows over-segmentation of SegMamba at screening and follow-up2 time points, over-segmentation of OmniMamba4D at follow-up1 and follow-up2 time points, and under-segmentation of OmniMamba at follow-up3 time point. For the lung lesions (third and fourth rows from Fig. 5), SegMamba and OmniMamba generate comparable segmentation results. In the first example liver lesion (fifth row from Fig. 5), the cropped volume of follow-up2 time point does not have the ground truth segmentation due to misregistration error. While SegMamba model generates a false positive segmentation, OmniMamba4D model correctly identifies the region as background. However, the Dice score does not account for true negatives in the absence of ground truth associated, which limits its ability to fully reflect OmniMamba4D's superior performance in this case.

## 4.3 Effectiveness of Tetra Mamba modules

An ablation study was conducted to assess the effectiveness of the Tetra Mamba module. As shown in Table 4, the Tetra Mamba version of OmniMamba4D outperformed the tri-orientated Mamba (ToM) variant. The base model with Tetra Mamba achieved a test Dice score of 68.227, compared to 67.494 for the ToM. This demonstrates that the Tetra Mamba module improves segmentation accuracy by effectively modeling temporal features in longitudinal CT images. We further compared a base OmniMamba4D model with feature sizes of [48, 96, 192, 384, 768] and a large OmniMamba4D model with feature sizes of [64, 128, 256, 512, 1024] across the five levels. Increasing the feature map size in the large model did not show improvement. It could be due to the relatively small size of the training data.

**Table 4.** Ablation study for different modules of Mamba.
† denotes a large model.

| Methods | Core Modules | Param (M) | Validation Dice ↑ | Test Dice ↑ |
|---|---|---|---|---|
| OmniMamba4D | ToM | 67.36 | 66.629 | 67.494 |
| OmniMamba4D† | | 102.43 | 66.364 | 67.421 |
| OmniMamba4D | Tetra-Mamba | 67.36 | **66.669** | **68.227** |
| OmniMamba4D† | | 102.81 | 66.620 | 67.815 |

## 5. CONCLUSION

OmniMamba4D, with its Tetra Mamba module, demonstrates competitive segmentation performance on longitudinal CT data by effectively capturing both spatial and temporal features. Despite handling more complex and larger input sequences, it achieves results on par with 3D models while maintaining computational efficiency. For the 16 lesions, where ground truth does not exist due to lesion disappearance, SegMamba generated false positive segmentations for all of them while OmniMamba4D generated segmentation for 12 of them and calling 4 lesions disappeared correctly. OmniMamba4D model is better at segmentation of lesions that could potentially disappear in response to treatment. Further improvements such as including a classification head as a multi-task learning framework [24] could help 4D model achieve better performance in these challenging disappeared lesions.

**Compliance with Ethical Standards:** The images are anonymized, and the imaging protocols are reviewed and approved by the Institutional Review Board.

**Conflict of interest:** No funding was received for conducting this study. The authors have no relevant financial or non-financial interests to disclose.

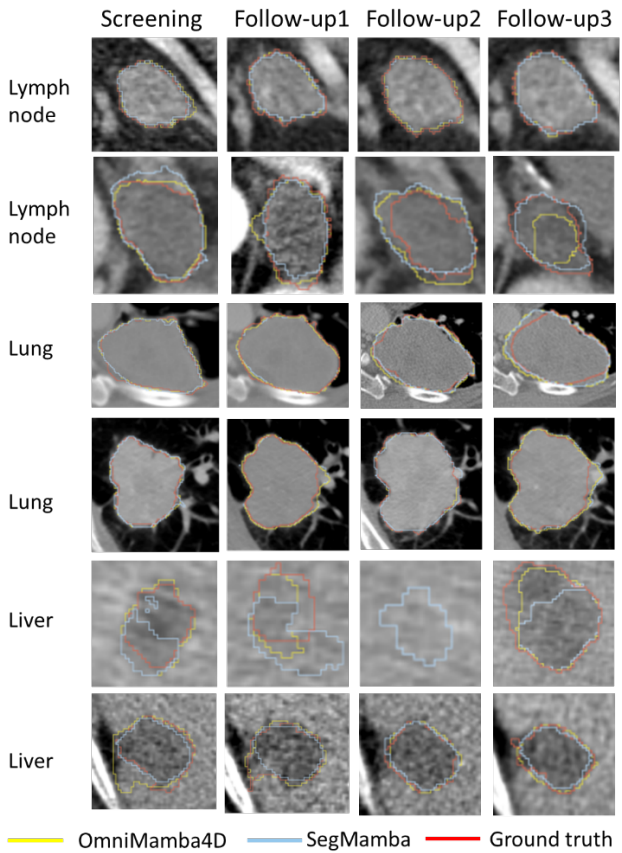

**Fig. 5.** Visualization of segmentation results comparing OmniMamba4D (yellow), SegMamba (blue), and ground truth (red). The figure shows six registered axial slices of lesions (rows) across four time points (columns).

## 6. REFERENCES

[1] L. C. Michaelis, M. Ratain *et al.*, "Measuring response in a post-RECIST world: from black and white to shades of grey," Nature Reviews Cancer, 6, 409-414, (2006).

[2] B. Song, K. Yang, J. Garneau *et al.*, "Radiomic Features Associated With HPV Status on Pretreatment Computed Tomography in Oropharyngeal Squamous Cell Carcinoma Inform Clinical Prognosis," Frontiers in Oncology, 11, (2021).

[3] M. L. Maitland, J. Wilkerson, S. Karovic *et al.*, "Enhanced Detection of Treatment Effects on Metastatic Colorectal

*Contributed as intern at Merck & Co., Inc., Rahway, NJ, USA